\documentclass{style}
\usepackage{graphicx}
\usepackage{epsfig}

\begin{document}
\begin{frontmatter}
\title{Coulomb distortion effects in quasi-elastic (e,e') scattering
on heavy nuclei}
\author{Andreas Aste}
\ead{andreas.aste@unibas.ch}
\address{Department of Physics, University of Basel, Klingelbergstrasse 82,
4056 Basel, Switzerland}
\date{March 7, 2008}
\begin{abstract}
The influence of the Coulomb distortion for quasi-elastic (e,e')
scattering on highly charged nuclei is investigated in distorted wave
Born approximation for electrons. The Dirac equation is solved numerically
in order to obtain exact electron continuum states in the electrostatic field
generated by the charge distribution of an atomic nucleus.
Different approximate models are used to describe the nucleon current
in order to show that, at high electron energies and energy-momentum transfer,
the influence of Coulomb distortions on (e,e') cross sections can be
reliably described by the effective momentum approximation, irrespective
of details concerning the description of the nuclear current.
\end{abstract}
\end{frontmatter}

\maketitle

\section{Introduction}

Quasi-elastic electron scattering on nuclei has represented
during nearly three decades one of the most successful tools to study
the nuclear and nucleon structure. Experiments have been
performed at SLAC \cite{Baran88,Chen91,Meziani92}, the MIT Bates Laboratory
\cite{Altemus83,Deady83,Hotta84,Deady86,Blatchley86,Dytman88,Dow88,Yates93,Williamson97}, the Saclay Laboratory
\cite{Barreau83,Meziani84,Meziani85,Marchand85,Zghiche94,Gueye99},
and more recently at JLab in order to explore this reaction.
Both inclusive $(e,e')$
and exclusive (single arm $(e,e'N)$ or double arm
$(e,e'NN)$) experiments contributed to a deeper understanding of the
many-body structure of strongly interacting systems like light
and heavier nuclei opening the possibility of investigating 
also the {\it in-medium} nucleon properties.

Inclusive $(e,e')$ scattering, where only the scattered electron is observed,
provides information on a number of
interesting nuclear properties like, e.g., the nuclear Fermi momentum
\cite{Whitney74}, high-momentum components in nuclear wave
functions \cite{Benhar94a}, modifications of nucleon form factors
in the nuclear medium \cite{Jourdan96a,MorgenMeziani}, the scaling properties
of the quasi-elastic response allow to study the reaction
mechanism \cite{Day90}, and extrapolation of the quasi-elastic
response to infinite nucleon number
$A=\infty$ provides us with a very valuable observable
of infinite nuclear matter \cite{Day89}.

However, experimental studies of inclusive and exclusive reactions
induced by electrons are hampered in the case of target 
nuclei with a large number of protons by the strong Coulomb 
field, which induces a distortion of the electron wave front, hereby modifying 
the structure of the $(e,e')$ cross section and inducing
sizable effects in the longitudinal and transverse separation
of the electromagnetic response \cite{Uber,CoHei87,TrTu87,TrTuZg88,Rho04}.

Still, there is a considerable theoretical and experimental interest 
in extracting longitudinal and transverse structure functions as a function 
of energy loss for fixed three momentum transfer for a range of 
nuclei. One of the important topics mentioned above concerning the
in-medium modification of the nucleon form factors is related
to the question whether the Coulomb sume rule is violated in nuclei.
This sum rule states that the number of protons in a nucleus can be
obtained from an integral of the electric response of the nucleus over
the full range of the electron energy loss at large three-momentum
transfer \cite{MorgenMeziani}.
Unfortunately, the conclusions reached by the large experimental programs
carried out at Bates, Saclay and SLAC for a variety of nuclei ranged from a
full saturation of the Coulomb sum rule to its violation by $30 \%$.

The approval of the Thomas Jefferson National Accelerator Facility
(TJNAF) Proposal E01-016, entitled `Precision measurement of longitudinal
and transverse response functions of quasi-elastic electron scattering in
the momentum transfer range $0.55$ GeV $\le$ $|\vec{q}|$ $\le$ $1.0$ GeV',
resulted in experiments performed very recently in Hall A at TJNAF, using
target nuclei like $^4$He, $^{12}$C, $^{56}$Fe, and $^{208}$Pb.
In the present study, we therefore focus on the particular case of
$^{208}\mbox{Pb}(e,e')$ scattering. Preliminary calculations for
$^{56}\mbox{Fe}(e,e')$ show that the conclusions
drawn below in the present study apply to heavy nuclei like $^{56}$Fe in an
analogous manner as well.

Since the plane wave Born approximation (PWBA) for electrons
is no longer adequate for the calculation of scattering
cross sections in the strong and long-range electrostatic field of
highly charged nuclei, it has become clear
in recent years that the correct treatment
of the Coulomb distortion of the electron wave function due
to the electrostatic field of the nucleus is unavoidable
if one aims at a consistent interpretation of experimental data.
The theoretical framework to investigate Coulomb corrections 
to the electron-nucleus cross sections is well 
established and is called distorted wave 
Born approximation (DWBA) in contrast to the better known
PWBA where the incoming and 
outgoing charged leptons are described by (Dirac) plane waves,
neglecting the effect of the Coulomb interaction between the
projectile and the target nucleus.
The application of the DWBA scheme is in principle straightforward,
but the numerical complications lead to extremely time consuming
calculations.

DWBA calculations with exact Dirac wave functions
have been performed by Kim {\emph{et al.}} \cite{Jin1} in the Ohio group
and Udias {\emph{et al.}} \cite{Udias,Udias2} for inclusive quasi-elastic
scattering on heavy nuclei. However, these cumbersome calculations are
difficult to control by people who were not directly involved
in the development of the respective programs. Early DWBA calculations for
$^{12}$C and $^{40}$Ca were presented in \cite{CoHei87}.
Coulomb corrections have also been evaluated theoretically by a group from
Trento University \cite{TrainiBeyond}, where it was found that the
standard method (used mainly in the case of light nuclei) to handle Coulomb
distortions for elastic scattering in data analysis, the so-called effective
momentum approximation (EMA), works with an accuracy better than
$1 \%$ for the description of Coulomb distortion effects, if the energy of
the scattered electrons and the momentum transfer is sufficiently high.
On the other hand, the Ohio group derived significant corrections beyond
the EMA. The findings presented in this paper confirm that the conclusions
drawn by the Trento group are basically correct.
There is a certain mismatch between the EMA and exact calculations for small
energy transfer, which, however, strongly decreases when the energy transfer
becomes significantly larger than the typical removal energy of the bound
nucleons.

Calculations using an eikonal approximation (called eikonal distorted
wave Born approximation, EDWBA) for electrons have been performed
by the Basel group \cite{Basel1,Basel2}, which seemed to confirm the
results of the Ohio group.
However, a poor approximation for the focusing of the
electrons near the nucleus was used, leading to an overestimation of
the Coulomb corrected cross sections. Through a subsequent analysis using
exact electron wave functions we found that the eikonal approximation with
an improved non-perturbative description of the electron wave function
amplitude supports the observation that the EMA is indeed a valid tool for
the description of Coulomb distortion effects \cite{SaclayPeak}.
A similar preliminary result was obtained in \cite{Basel3}, where exact
solutions of the Dirac equation were used for the electron wave functions.
The nuclear current was modeled within the framework of a simplified single
particle shell (SPS) model, with harmonic oscillator wave functions
for the bound nucleons and plane waves for the knocked-out nucleons.
The cross section for inclusive quasi-elastic scattering process
is usually calculated approximately by summing over all knock-out
processes, where the individual bound nucleons inside the nucleus
leave the nucleus after a sufficiently high energy transfer
$\omega=\epsilon_i-\epsilon_f$ from the scattered electron, where
$\epsilon_{i}$ is the initial and $\epsilon_{f}$
the final electron energy.
The recoil nucleons move in an energy dependent
optical potential. It is common practice
that the imaginary part of the optical potential is not taken
into account in the calculations, since the imaginary part is intended to
describe the loss of nucleon flux inside the nuclear medium. In inclusive
processes, only the electrons are observed, and one may argue that it is
not necessary to take into account whether or not the recoil
nucleons actually leave the nucleus or initiate some subsequent
nuclear reactions. Still, it has been shown in \cite{Horikawa}
that $(e,e')$ peak intensities are reduced by typically $5-10$\%
as a result of final state interaction (FSI) for the kinematics
relevant for this paper, and how this reduction is related
to the imaginary part of the final state nucleon optical potential
(see eq. (29a) in \cite{Horikawa}). The effects of the FSI
at high momentum transfer have been discussed in \cite{Benhar91}.

In this paper, we present calculations based on two different models
for the nuclear current. The final state nucleons are not described by
wave functions obtained as solutions of the Dirac equation with some
nuclear model potential.
Instead, an eikonal approximation and the plane wave approximation are
used, reducing the computational costs in a very effective way.
It turns out that the use of approximate wave functions is not a weakness,
since we find that at sufficiently high energies and momentum transfer,
the effective momentum approximation provides an accurate description
of Coulomb corrections, irrespective of the nuclear model used. It is a
fact that also elaborate nuclear current models, based, e.g., on
solutions of the Dirac equation for the nucleons in some relativistic
nuclear model, cannot be considered as a fully satisfying strategy,
as the still sizeable discrepancy between measured and calculated
cross sections shows (see, e.g., \cite{Jin1}).
Computational costs always restrain us from using
more realistic nuclear models, where effects like nucleon correlations and
meson exchange currents can be properly taken into account.
The electrons are described by exact solutions of the Dirac equation in our
DWBA calculations.

\section{From DWBA to EMA}
Inclusive $(e,e')$ cross sections are usually calculated by integrating
over all final state nucleons in the knock-out $(e,e'N)$ cross sections
obtained from a single-particle shell model \cite{Goeppert} for the
bound protons and neutrons,
as will be explained in further detail in the following section.
The differential cross section for single nucleon knockout is
given by \cite{Udias}
\begin{displaymath}
\frac{d^4 \sigma}{d \epsilon_f d\Omega_f dE_f d\Omega_f}=
\end{displaymath}
\begin{equation}
\frac{4 \alpha^2}{(2 \pi)^3} \epsilon_f^2 E_f p_f
\delta(\epsilon_i \! + \! E_A \! - \! \epsilon_f \!- \! E_f
\! - \! E_{A-1}) \sum \limits_{}^{ - \! \! -} \, |W_{if}|^2
\label{crosssection},
\end{equation}
with the matrix element
\begin{equation}
W_{if}=\frac{1}{(2 \pi)^3}
\int \! d^3 r_e \!  \int d^3 r_N \! \int \! d^3 q' \Bigl[ j_\mu^e(\vec{r}_e) \,
\frac{e^{-i\vec{q}{\, '} (\vec{r}_e-\vec{r}_N)}}{{q'}_\mu^2+i0} \,
j_N^\mu(\vec{r}_N) \Bigr], \label{matrixelement}
\end{equation}
where $j_N^\mu(\vec{r}_N)$ is the nucleon transition current obtained
within the framework of some suitable nuclear model,
the $\sum \limits_{}^{ - \! \! -}$ in eq. (\ref{crosssection})
indicates the sum (average) over final (initial)
polarizations, and $E_A$, $E_{A-1}$ is the energy of the initial and
final nucleus, respectively. The virtuality of the exchanged photon
is ${q'}_\mu^2=\omega^2-\vec{q}{\, '}^2$, where $q^0=\omega=\epsilon_i-
\epsilon_f$ is the energy tranferred by the electron to the nuclear system.
In our calculations, the final nucleus was considered as a spectator, and the
final energy $E_f$ of the nucleon was calculated from the energy transfer
reduced by a reasonable separation (or removal) energy.

In the PWBA, the electron current is given by
\begin{equation}
j_e^\mu(\vec{r}_e)=
\bar{u}_{s_f}(\vec{k}_f) \gamma^\mu u_{s_i}(\vec{k}_i)
e^{i \vec{k}_i \vec{r}_e-i \vec{k}_f \vec{r}_e},
\end{equation}
where ${u}_{s_i},{u}_{s_f}$ are initial/final state
plane wave electron spinors corresponding to the initial/final electron
momentum $\vec{k}_{i,f}$ and spin $s_{i,f}$, normalized according to
\begin{equation}
\bar{u}_s(\vec{k}) \gamma^0 u_{s'}(\vec{k})=\delta_{s,s'},
\end{equation}
and a short calculation
yields a three-dimensional integral
\begin{equation}
W_{if}^{PWBA} = - \frac{1}{Q^2} \int d^3 r
\Bigl[ j_\mu^e(\vec{r}) j_N^\mu(\vec{r})\Bigl].
\end{equation}
with the four-momentum transfer squared $Q^2=(\vec{k}_i-\vec{k}_f)^2-\omega^2=
\vec{q}^{\, 2}-\omega^2$.

In the DWBA, exact solutions of the Dirac equation
for electrons in the electrostatic field of the nucleus are used instead
of plane wave functions. In a general sense, the EMA accounts for the
two effects of the Coulomb distortion, namely the momentum enhancement
of the electron near the attractive nucleus and the focusing of the
electron wave function.
Therefore, transition amplitudes are calculated by replacing the
asymptotic initial and final state momenta $\vec{k}_i$, $\vec{k}_f$
of the electron by appropriate effective momenta in the PWBA
transition amplitude. Additionally, the cross section
obtained from the effective transition amplitude is multiplied by
an appropriate focusing factor, which accounts for the fact that
the electron wave function amplitude is enhanced in the vicinity of
the nucleus, i.e. from a semiclassical point of view,
the nucleus acts like a lens and concentrates the electrons towards the
nuclear interior.

Indeed, one may establish a direct connection between the
DWBA and the EMA. The DWBA transition amplitude comprises the exact
solution of the Dirac equation for the electron in the electrostatic field
of the nucleus. In this way, the DWBA accounts for the exchange of the `soft'
photons between electron and nucleus, whereas the explicit photon propagator
in eq. (\ref{matrixelement}) accounts for the exchange of one hard photon.

Using the distributional identity
\begin{equation}
\int \frac{d^3 q'}{(2 \pi)^3}
\frac{e^{i \vec{q}{\, '} \vec{r}}}{\omega^2-\vec{q'}^2 \pm i0} =
- \frac{e^{\pm i \omega r}}{4 \pi r},
\end{equation}
one obtains
\begin{displaymath}
W_{if}=-\frac{1}{4 \pi} \int d^3 r_e d^3 r_N \, j_e^\mu ((\vec{r}_e)
j^N_{\mu} (\vec{r}_N)
\frac{e^{i \omega |\vec{r}_e-\vec{r}_N|}}{|\vec{r}_e-\vec{r}_N|}=
\end{displaymath}
\begin{displaymath}
-\frac{1}{4 \pi} 
\int d^3 r_e d^3 r_N \Bigl\{ \rho_e (\vec{r}_e) \rho_N (\vec{r}_N)-
\vec{j}_e(\vec{r}_e) \vec{j}_N (\vec{r}_N) \Bigr\}
\frac{e^{i \omega r}}{r},
\end{displaymath}
\begin{equation}
\quad r=|\vec{r}_e-\vec{r}_N|,
\end{equation}
with $\rho_e, \, \rho_N, \, \vec{j}_e, \, \vec{j}_N$ denoting the transition charge
and current densities of the electron and the nucleon, respectively.
The double volume integral presents a clear numerical disadvantage of this expression.
According to Knoll \cite{Knoll1}, one may introduce the scalar operator
\begin{equation}
S=e^{i \vec{q}\vec{r}} \sum \limits_{n=0} \Biggl(
\frac{2i \vec{q} \, \vec{\nabla} +\Delta}{\vec{q}^{\, 2}-\omega^2} \Biggr)^n
e^{-i \vec{q} \vec{r}}, \quad \vec{q}=\vec{k}_i-\vec{k}_f, \label{scalarop}
\end{equation}
such that the transition amplitude can be expanded in a more convenient
form
\begin{equation}
W_{if}= -\frac{1}{Q^2} \int d^3 r
\Bigl[ \rho_N (\vec{r}) S \rho_e (\vec{r}) -
\vec{j}_N S \vec{j}_e (\vec{r}) \Bigr]. \label{wif}
\end{equation}
The single integral is limited to the region of the nucleus, where the
nuclear current is relevant. The expansion eq. (\ref{scalarop})
is an asymptotic one, in the sense that there is an optimum number
(depending on $\vec{q} \, $) of terms that give the best approximation
to the exact value.
Considering terms up to second order in the derivatives only one obtains
\begin{displaymath}
W_{if} =-\frac{1}{Q^2} \int d^3 r
\Biggl\{
\rho_N (\vec{r}) e^{i \vec{q} \vec{r}} \Biggl[
1 + \frac{2 i \vec{q} \, \vec{\nabla} + \Delta}{Q^2}
-\frac{4 ( \vec{q} \,  \vec{\nabla})^2}{(Q^2)^2} \Biggr]
e^{-i \vec{q} \vec{r}} \rho_e(\vec{r})
\end{displaymath}
\begin{equation}
+ \mbox{current terms} \Biggr\}. \label{approx}
\end{equation}
This approximation has been used in our calculations \cite{Basel3}
and in an equivalent way in \cite{Jin1},
where it was also observed that higher order terms can indeed be
neglected for the calculations relevant for this work.
It should also be mentioned that the approximation has been
applied for the first time to inclusive reactions in
\cite{TrTu87,TrTuZg88}, and discussed in the context
of the $(e,e'p)$ quasi-elastic reaction in \cite{GiuPa87}.

There is indeed a simple semiclassical interpretation of this expansion.
If one assumes that the electron is scattered locally with a momentum
transfer $\vec{q}_{loc}(\vec{r})=\vec{q}+\Delta \vec{q}(\vec{r})$,
calculable from the classical local
momentum of the electron in the electrostatic potential and which deviates by $\Delta \vec{q}$
from the the asymptotic momentum transfer $\vec{q}$, then one has
\begin{displaymath}
\frac{Q^2}{Q_{loc}^2}=\frac{\vec{q}^{\, 2}-\omega^2}{(\vec{q}+\Delta{\vec{q}} \, )^2-\omega^2}=
\frac{Q^2}{Q^2+2 \vec{q} \Delta \vec{q} + \Delta \vec{q}^{\, 2}}=
\end{displaymath}
\begin{equation}
1-\frac{2 \vec{q} \Delta \vec{q} + \Delta \vec{q}^{\, 2}}{Q^2} +
\frac{4 (\vec{q} \Delta \vec{q} \, )^2}{(Q^2)^2}+... \label{ClassKnoll}
\end{equation}
Furthermore, one may note that the PWBA electron transition current can be written as
\begin{equation}
j_e^\mu(\vec{r})=|j_e^\mu(\vec{r})| e^{i \vec{q} \vec{r}} .
\end{equation}
Applying the operator $\hat{\vec{q}}=\frac{\vec{\nabla}}{i}$ on $j_e^\mu(\vec{r})$
leads to $\hat{\vec{q}} j_e^\mu(\vec{r})=\vec{q} j_e^\mu(\vec{r})$,
i.e. $\frac{\nabla}{i}$ acts as momentum transfer operator and replacing $\vec{q}$
in (\ref{ClassKnoll}) by $\hat{\vec{q}}$, one immediately obtains the
differential operator part of $S$. 

Therefore, the Knoll operator $S$ first factors out the undistorted
$e^{i \vec{q} \vec{r}}$-term in the current, then calculates the ratio
of the asymptotic momentum transfer with a quantity that can be
interpreted as the local momentum transfer generated from the
distorted electron current, and finally reinserts the $e^{-i \vec{q} \vec{r}}$-behavior
of the undistorted current.

Assuming that one can approximate the effect of the $S$-operator by replacing it
by an average (effective) momentum transfer $Q_{eff}^2$, one obtains as a first approximation
for the transition matrix element
\begin{equation}
W_{if} =-\frac{1}{{Q}_{eff}^2} \int d^3 r j_e^\mu (\vec{r}) j^N_\mu (\vec{r}), \label{appro1}
\end{equation}
keeping in mind that the DWBA electron current still has to be used in
eq. (\ref{appro1}).
Since a highly relativistic particle (with negligible mass $m$) is
moving nearly on a straight line,
the classical momentum of a particle with asymptotic momentum
$\vec{k}$ and energy $\sqrt{\vec{k}^2+m^2}\simeq |\vec{k}|$ moving
inside a potential $V(\vec{r})$ is given by $\frac{\vec{k}}{|\vec{k}|}(k-V(\vec{r}))$.
Additionally, since the knockout process is nearly local for large
momentum transfer (e.g., $Q^2=(400 \, \mbox{MeV})^2$ corresponds to
a photon propagation length scale of $\simeq \frac{1}{2} \, \mbox{fm}$
which is much smaller than the extension of a heavy
nucleus), and since it takes place in the entire volume of the nucleus,
it is self-evident that one should calculate ${Q}_{eff}^2$ from a
potential value $\bar{V}$ which is obtained by an averaging process
over the nuclear density profile $\rho(\vec{r})$
\begin{equation}
{Q}_{eff}^2=\Bigl( \vec{k}_i^{eff}-\vec{k}_f^{eff} \Bigr)^2 - \omega^2, \quad
\vec{k}_{i,f}^{eff}= \frac{\vec{k}_{i,f}}{|\vec{k}_{i,f}|}(k_{i,f}-\bar{V})
\end{equation}
with
\begin{equation}
\bar{V} \simeq \frac{\int d^3 r \rho(\vec{r}) V(\vec{r})}{\int d^3 r
\rho (\vec{r})},
\end{equation}
where $\rho$ can also be used for an approximate description of the
charge density of the nucleus.
In the case of a homogeneously charged sphere with radius $R$ and
charge number $Z$, the electric potential in the center of the sphere
is given by $V(0)=-\frac{3 \alpha}{2 R}$
($\alpha$ is the fine structure constant), whereas the potential
averaged over the volume of the sphere is given by $\bar{V}=4 V(0)/5$.

A further approximation which can be made in eq. (\ref{appro1}) stems
from the observation that the attractive nucleus focuses the electron
wave function in the nuclear region.
It can be shown that for highly relativistic energies, the amplitude
of the initial/final state electron wave function in the nuclear
center is enhanced by a focusing factor
$f_{i,f}(0)=(1-V(0)/|\vec{k}_{i,f}|)$, however, solving the Dirac
equation for an electron in the electrostatic field of a highly
charged nucleus again shows that the focusing averaged over
the nuclear volume is again well described by $f_{i,f}=(1-\bar{V}/|\vec{k}_{i,f}|)$.
Since the electron transition current is given by $j_e^\mu = \bar{\psi}_{e,f} \gamma^\mu
\psi_{e,i}$, one may replace $j_e^\mu(\vec{r})$ by $f_i^{eff}
f_f^{eff} j_e^{\mu,EMA}$,
where $j_e^{\mu,EMA}$ is calculated with plane electron waves with
effective momenta.
A detailed discussion of the focusing effect can be found in \cite{Aste2006}.

If the same average potential value $\bar{V}$ is used to calculate the
effective momenta and the effective focusing factors, then a short calculation shows that
the focusing factors cancel against the effective photon propagator
$\sim Q^{-4}_{eff}$ in the cross section in the sense that
\begin{equation}
\frac{(f_i^{eff})^2 (f_f^{eff})^2}{Q^4_{eff}}=\frac{1}{Q^4},
\end{equation}
and consequently, the EMA cross section is calculated by replacing the electron current
in (\ref{approx}) by the PWBA current calculated from the effective momenta, and leaving
everything else unchanged. A different, but completely equivalent strategy is to calculate first
the theoretical cross section based on the effective momenta and energies instead of the
corresponding asymptotic values. The intermediate result obtained this way must be
multiplied subsequently by $(f_i^{eff})^2$ in order to account for the focusing of the
incident electron. The focusing of the final state electron is automatically taken into
account by the artificially enhanced phase space factor
$(\epsilon_f^{eff})^2 \sim |\vec{k}_f^{eff}|^2$.

The cancellation mechanism that plays between the focusing effect and the the photon propagator
in the matrix element is accidental and not really exact. It works in a
very satisfactory way for nuclear charge distributions which are close
to the homogeneous, spherical case. If the energy of the final state
electron becomes significantly smaller than
$150 \, \mbox{MeV}$ and the four-momentum transfer is below
$(300 \, \mbox{MeV})^2$, the semiclassical picture on which the EMA
is based starts to fail. Furthermore, for small energy transfer
$\omega$ comparable to the removal energy of the nucleons, details of
the nuclear structure and the interaction of the recoil nucleons with
the nuclear matter become increasingly important,
which makes the applicability of the EMA questionable. However, this 
kinematical region is of minor interest in future experiments.

\section{Models and approximations}

\subsection{Single particle shell model}
As mentioned before, inclusive $(e,e')$ cross sections are commonly
calculated by integrating over all final state nucleons in $(e,e'N)$
cross sections obtained from a single-particle shell (SPS) model for
the bound protons and neutrons.
Usually, the fact that there are short-range and tensor correlations
between nucleon pairs which lead to a partial depletion of the single particle shells
is ignored in SPS $(e,e')$ calculations.
Within a correlated basis functions theory calculation, it was found
in \cite{Benhar1989} for nuclear matter that there is an approximate
probability of 20\% for a nucleon being in an correlated state
\cite{Sick2007}. However, correlated nucleons account for
approximately 37\% of the average removal energy. In view of the fact
that the average binding energy of, e.g., protons is $21$ MeV in our
SPS calculations, this observation implies that the correlated
nucleons correspond to strongly bound particles with binding energies
around $100$ MeV, if an average removal energy of $37$ MeV is
assumed.
The impact of correlated nucleons on the inclusive cross
section is therefore strongly suppressed for an energy transfer
clearly smaller than $150$ MeV compared to the strength of the
uncorrelated nucleons. Still, a reduction of $\sim 20$\% of the
peak strength results from the suppression of correlated nucleons,
combined with a slight shift of the peak towards higher energy
transfer $\omega$ and an increase of strength in the
high-$\omega$ tail.

A further observation, which is of minor importance concerning
the present calculations, is the fact that the rms radius of
uncorrelated nucleons tends to be slightly larger
than the total rms radius of the nucleus \cite{Sick2007}.
We therefore performed calculations within a strongly simplified
framework, where the correlated (high-momentum) nucleons where described
by model wave functions with a high-momentum component and
large binding energies. We found that even in this case, the EMA
remains valid to an acceptable extent, if the `correlated' nucleons
are distributed in a reasonable homogeneous manner inside
the nucleus, and if the momentum transfer and final electron energy
are sufficiently large.
In this work, we therefore focus on the main contribution of the
uncorrelated nucleons in the region of the quasi-elastic peak,
in order to facilitate a comparison to other works \cite{Jin1}.

The relativistic form of the wave functions for bound protons and neutrons
was constructed from the non-relativistic wave function.
Non-relativistic bound nucleon wave functions were generated using a
self-consistent Schr\"odinger solver for a non-relativistic Woods-Saxon
potential and the additional Coulomb potential for the protons.
The Woods-Saxon potential, which included an LS coupling term, was
optimized in such a way that the experimental binding energies
of the upper proton shells and the rms of the
nuclear charge distribution were reproduced correctly.
The standard way to construct nucleon four-spinors $\psi_N$ from
a Schr\"odinger-type nucleon wave function $\chi_N$ was then
applied, given by
\begin{equation}
\psi_N (\vec{x}) =
N \Bigl(\chi_N (\vec{x}), \frac{\vec{\sigma} \cdot
\hat{\vec{p}}}{\bar{E}+m_N} \chi_N(\vec{x}) \Bigr),
\end{equation}
where  $\hat{\vec{p}}$ denotes the differential momentum operator,
$\bar{E}$ is the energy of the nucleon,
and $N$ is a normalization factor which is typically close to one.
The same strategy, which is exact in the case of Dirac plane waves,
was also applied to construct the small component of
the Dirac spinor in the case of the nucleon eikonal approximation.
For a discussion of this approximation we refer to \cite{Amaro}.

The nuclear charge distribution, which is the main input for the calculation
of the Dirac electron wave functions, and nucleon optical potentials of heavy nuclei are often
approximated by the help of a Woods-Saxon distribution.
Since the normalization and moments of Woods-Saxon distributions
are rarely found in the literature, some important expressions are listed in
Appendix A.

\subsection{Eikonal approximation for nucleons}

In order to generate realistic nucleon wave functions in a
computationally efficient way, we used eikonal wave functions for the
final state protons and neutrons, with the aim to compare the
(Coulomb corrected) cross sections obtained from the eikonal nuclear
model to the cross section obtained from simple nucleon plane
wave calculations. This comparison will serve as a test
whether the applicability of the EMA is strongly dependent
on the nuclear model or not.

The eikonal phases $\chi_{1,2}$ generated by a potential $U(\vec{r})$
take the local (semiclassical) momentum modification of
the incoming/outgoing nucleon with initial/final asymptotic
momentum $\vec{p}_{i,f}=| {\vec{p}}_{i,f} | \hat{p}_{i,f}$ approximately
into account by modifying the plane waves describing the nucleons according to
\begin{equation}
e^{i\vec{p}_{i,f}\vec{r}} \, \rightarrow \, e^{i \vec{p}_{i,f}
\vec{r} \pm i\chi_{1,2}(\vec{r})} \, ,
\end{equation}
where
\begin{equation}
\chi_1(\vec{r})=-\frac{E_i}{|\vec{p}_i|}
\int \limits_{-\infty}^{0} U(\vec{r}+ \hat{p}_i s) ds
\, , \quad z=\hat{p}_i \vec{r} \, ,
\end{equation}
or
\begin{equation}
\chi_1(\vec{r})=-\frac{E_i}{|\vec{p}_i|} \int \limits_{-\infty}^{z}
U(x,y,z') dz' \, , \quad z=\hat{p}_i \vec{r} \,
\end{equation}
if the initial nucleon momentum is parallel to the $z$-axis.
The eikonal phase for the final state relevant for the calculations
presented in this work is
\begin{equation}
\chi_2(\vec{r})=- \frac{E_f}{|\vec{p}_f|} \int \limits_{0}^{\infty}
U(\vec{r}+ \hat{p}_f s') ds' \, ,
\end{equation}
where $E_{i,f}=({\vec{p}}_{i,f}^{\, 2}+m_N^2)^{1/2}$ is the energy of
the nucleon with mass $m_N$, and $E_f/|\vec{p}_f|$ is the velocity of the
knocked-out nucleon.
An energy-dependent volume-central part of an
optical Woods-Saxon type model potential $V_{nuc}(r,E)$
as given in a recent work \cite{Koning}
was used for our calculations, as well as the energy-independent
Woods-Saxon potential which was used to generate the bound state
nucleon wave functions.

The total potential $U(r,E)=V_{nuc}(r,E)+V_{coul}(r)$ is given as the
sum of a Woods-Saxon potential and
the Coulomb potential of the final state nucleus in the case of protons.
The depth of the Woods-Saxon type nuclear potential
\begin{equation}
V_{nuc}(r,E)=-\frac{V_{WS}(E)}{1+e^{(r-r_{nuc})/a_{nuc}}}
\end{equation}
depending on the energy $E$ of the proton (all quantities in MeV) is given by
\begin{equation}
V_{WS}(E)=v_1 [ 1- v_2 (E-E_f)+v_3 (E-E_f)^2-v_4(E-E_f)^3 ] \, ,
\end{equation}
where $E_f=-5.9$ and
\begin{equation}
v_{1}=67.2 , \, v_2=7.9 \cdot 10^{-3}, \,
v_3=2.0 \cdot 10^{-5}, \, v_4=7 \cdot 10^{-9},
\end{equation}
whereas the parameters for neutrons are $E_f=-5.65$ and
\begin{equation}
v_{1}=50.6, \, v_2=6.9 \cdot 10^{-3}, \,
v_3=1.5 \cdot 10^{-5}, \, v_4=7 \cdot 10^{-9}.
\end{equation}
The range and diffusivity are both for protons and neutrons given by
$r_{nuc}=7.371 \,$ fm and $a_{nuc}=0.646 \,$ fm according to \cite{Koning}.

The imaginary part of the optical potential, which is intended to
describe the loss of flux in proton (neutron) elastic scattering,
has been neglected in our $(e,e')$ calculations.
As mentioned in the introduction, in inclusive processes, only the
electrons are observed, and the eventuality whether or not an
ejected nucleons got `lost', i.e. initiated some subsequent nuclear
reaction, does not play a very important role. Furthermore, the FSI
affects the inclusive cross sections in a similar way
whether the electron wave function distortion is taken into account or
not, such that a comparison of electron PWBA and DWBA calculations
is still possible.

\begin{figure}
        \centering
        \includegraphics[width=2.90in]{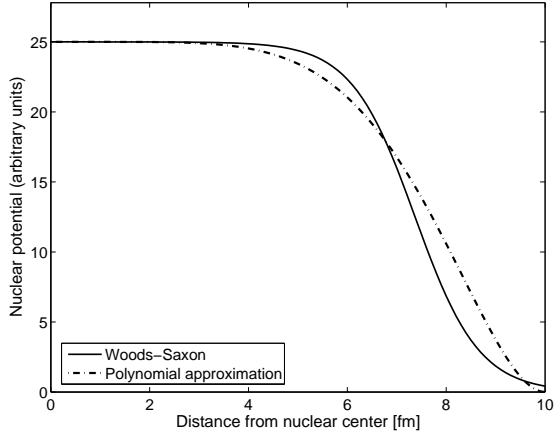}
        \caption{Comparison of the Woods-Saxon type nuclear potential and the simple
                 polynomial approximation described in the text, which allows to derive
                 a simple analytic expression for the nuclear eikonal phase.}
        \label{approxws}
\end{figure}

To calculate the eikonal integral
for a Woods-Saxon potential numerically would be relatively
time consuming, and unfortunately, the eikonal integral of a
Woods-Saxon potential is not suitable for simple analytic
estimations \cite{Lukyanov,Shepard}.
Therefore, one may approximate the Woods-Saxon shape of the
mean-field optical nucleon potential $V_{nuc}$ in an effective way by
a power series in the distance from the nuclear center.
We found that the following simple two-parameter approximation
provides an efficient approximation for $V_{nuc}(r)$
in the range $0 < r < R_{max}$ in the case of $^{208}$Pb,
where $R_{max}=10 \,$fm:
\begin{equation}
{\tilde{V}}_{nuc}(r,E)=-V_{WS}(E)+c_1(E) r^7+c_2(E) r^6
\label{threeparam}
\end{equation}
with
\begin{equation}
c_1=-6 \frac{V_{WS}}{R_{max}^{7}}, \quad
c_2=7 \frac{V_{WS}}{R_{max}^{6}}.
\end{equation}
The approximate nuclear potential ${\tilde{V}}_{nuc}(r,E)$ satisfies the boundary conditions
${\tilde{V}}_{nuc}(R_{max},E)=0$ and $\frac{d}{dr}
{\tilde{V}}_{nuc}(R_{max},E)=0$, which are acceptable for a distance of
$10 \,$fm from the nuclear center.
A comparison of the exact Woods-Saxon distribution and approximation
eq. (\ref{threeparam}) is shown in Fig. {\ref{approxws}}.
The contribution of ${\tilde{V}}_{nuc}(r,E)$ to the eikonal phase of the
nucleons can be derived from the integrals
\begin{displaymath}
\int (z^2+b^2)^3 dz = \frac{1}{7} z^7 + \frac{3}{5} b^2 z^5+ b^4 z^3 +b^6 z \, ,
\end{displaymath}
\begin{displaymath}
\int \sqrt{z^2+b^2}^7 dz = \Biggl[\frac{93}{128} b^6 z + \frac{163}{192} b^4 z^3+
\frac{25}{48} b^2 z^5 + \frac{1}{8} z^7 \Biggr]
\end{displaymath}
\begin{equation}
\times \sqrt{z^2+b^2}+
\frac{35}{128} b^8 \log(z+\sqrt{z^2+b^2}).
\end{equation}

In order to be compatible with \cite{Koning}, we used the
electrostatic Coulomb potential generated by an homogeneously
charged sphere ($Z'=Z-1$, $R_C=1.220 A^{1/3}= 7.228 \, {\mbox{fm}}$)
\begin{equation} \label{identities}
V_{coul}(r)=\left\{ \begin{array}{llc} \frac{\alpha Z'}{R_C}
\Bigl( \frac{3}{2}-\frac{r^2}{2 R_C^2} \Bigr) & : & r \leq R_C \\
\frac{\alpha Z'}{r} & : & r \geq R_C
\end{array} \right.
\end{equation}
for protons. The eikonal phase can easily be constructed
from the expressions above, however one should note that the
long-range part of the Coulomb potential must be regularized
and one has ($r'^2=b^2+z'^2$,
$r^2=b^2+z^2$, $z=\hat{p}_i \vec{r}$)
\begin{displaymath}
\chi_1^{coul}(\vec{r})=-\alpha Z' \frac{E_i}{|\vec{p}_i|}
\lim_{R \to \infty} \Bigl[ \int \limits_{-\infty}^{z} \,
\Bigl( \frac{1}{r'}-
\frac{1}{\sqrt{r'^2+R^2}} \Bigr) dz' -\log(R/R_0) \Bigr]=
\end{displaymath}
\begin{equation}
-\alpha Z'\frac{E_i}{|\vec{p}_i|}
\log \Biggl( \frac{R_0(r+z)}{b^2} \Biggr).
\end{equation}
or
\begin{displaymath}
\chi_1^{coul}(\vec{r})=
\end{displaymath}
\begin{equation}
-\alpha Z' \frac{E_i}{|\vec{p}_i|}
\log \Biggl( \frac{R_0(r+z)}{r^2-z^2} \Biggr)
=\alpha Z'\frac{E_i}{|\vec{p}_i|}
\log (k_i r - \vec{k}_i \vec{r})+const.,
\end{equation}
which is defined up to a constant phase given by the free
parameter $R_0$. Analogously, $\chi_2^{coul}$ which is relevant
for our case is given in a space region where, e.g., $r > R_C$ and
$\hat{p}_f \vec{r}>0$, by
\begin{displaymath}
\chi_2^{coul}(\vec{r}) = -\alpha Z' \frac{E_f}{|\vec{p}_f|}
\log \Biggl( \frac{R'_0(r-{\tilde{z}})}{{\tilde{b}}^2} \Biggr)
\end{displaymath}
\begin{equation}
=\alpha Z' \frac{E_f}{|\vec{p}_f|} \log (k_f r + \vec{k}_f \vec{r})
\end{equation}
with ${\tilde{z}}=\hat{p}_f \vec{r}$ and ${\tilde{b}}^2=r^2-{\tilde{z}}^2$.

\begin{figure}
\begin{center}
  \centerline{
    \mbox{\includegraphics[width=2.90in]{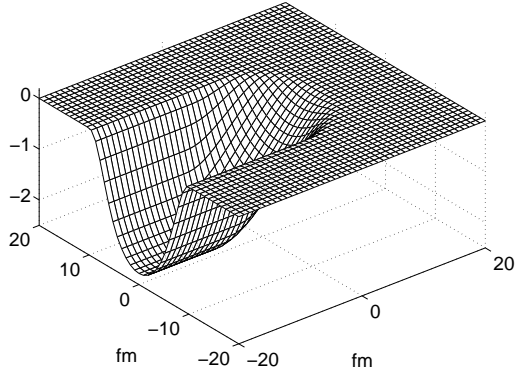}}
  }
        \caption{Typical eikonal phase for outgoing neutrons in a plane through the
                 nuclear center ($\vec{p}_f$ points rearwards to the right). The plot shows
                 $\int \limits_{0}^{\infty} V_{nuc}(\vec{r}+ \hat{p}_f s) ds$,
                 where $V_{nuc}(\vec{r})$ is an attractive Woods-Saxon potential with a depth given via
                 $V_{WS}(E)=30$ MeV and corresponding parameters for $^{208}$Pb.}
        \label{eiko_prot_n}
\end{center}
\end{figure}

\begin{figure}
\begin{center}
  \centerline{
    \mbox{\includegraphics[width=2.90in]{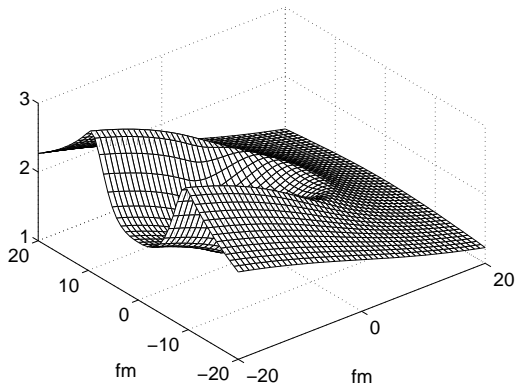}}
  }
        \caption{Typical eikonal phase for outgoing protons in a plane through the
                 nuclear center. The plot shows the regularized combined phase
                 $\mathcal{R} \int \limits_{0}^{\infty} U(\vec{r}+ \hat{p}_f s) ds$ with
                 $U(r)=V_{nuc}(r,E)+V_{coul}(r)$
                 for the same nuclear Woods-Saxon potential as in Fig. \ref{eiko_prot_n} 
                 and the repulsive electrostatic potential for $^{208}$Pb
                 with $Z'=81$.}
        \label{eiko_prot_p}
\end{center}
\end{figure}

Figs. \ref{eiko_prot_n} and \ref{eiko_prot_p}
show typical eikonal phases for neutrons (protons) in (a repulsive Coulomb and) a
mean field optical potential for $^{208}$Pb with a depth of, e.g., $30$ MeV.
Note that the maximal depth of the surface in Fig. \ref{eiko_prot_n} is given approximately by
the spatial extension of the Woods-Saxon potential multiplied by the potential depth,
i.e. $\sim 30 \, \mbox{MeV} \times 14 \, \mbox{fm} \simeq 2.13 \, \hbar c$.
The eikonal phases depicted in Figs. \ref{eiko_prot_n} and \ref{eiko_prot_p}
must be multiplied, up to a sign, by a energy-dependent factor $E_f/|\vec{p}_f | >1$ in order to get 
the relevant phase which distorts the corresponding nucleon Dirac spinor.
 
In addition to the eikonal phase correction, we also took the
(de-)focusing effect of the nucleon wave functions in the energy-dependent
optical potential into account.
We remark that Baker investigated the second-order eikonal approximation
for potential scattering in the non-relativistic case \cite{Baker},
finding thereby an expression for the focusing factor of continuum Schr\"odinger wave
functions. For a spherically symmetric potential $V(r)$, one finds for the central
focusing factor $f(0)$ of the amplitude of a non-relativistic scalar nucleon
wave function (see eq. (23) in \cite{Baker})
\begin{equation}
f(0) \simeq 1-\frac{V(0)}{2 p_f v_f},
\end{equation}
where $p_f$ is the asymptotic momentum and $v_f$ the velocity of the particle.
Roughly speaking, the approximation is valid if the asymptotic kinetic energy of the
particle is larger than the depth of the disturbing potential
$m \gg E^f_{kin}=E_f-m \gg |V(0)|$, and the wave length of the particle $\sim 2 \pi/p_f$
should be significantly smaller than the extension of the potential.
For the classical particle momentum in the center of the potential
$p_f(0)$ one has non-relativistically
\begin{displaymath}
p_f(0) = \sqrt{2m(E^f_{kin}-V(0))}
\end{displaymath}
\begin{equation}
=\sqrt{2 m E^f_{kin} ({1-V(0)/E^f_{kin}})}
\simeq p_f \Bigl( 1 - V(0)/2 E^f_{kin} \Bigr),
\end{equation}
such that
\begin{equation}
f(0) \simeq 1 - \frac{V(0)}{2 p_f v_f} \simeq \sqrt{\frac{p_f(0)}{p_f}},
\end{equation}
i.e. it is found that the probability density is enhanced by the square root of the ratio of the
central and asymptotic momenta $p_f(0)/p_f$, contrary to the result $f(0) = p_f(0)/p_f \simeq
(E_f-V(0))/E_f$ in the highly relativistic case.
One may ask how the non-relativistic and the highly relativistic regime
are connected. A classical relativistic analysis of the particle trajectories
shows that the central focusing is given by the expression
\begin{equation}
f(0)=\Biggl (\frac{p_f(0)}{p_f} \frac{E_f-V(0)}{E_f} \Biggr)^{1/2},
\end{equation}
which interpolates between the non-relativistic and relativistic regime \cite{Aste2007}.
We additionally solved the Dirac equation for Dirac nucleons in the
energy-dependent volume-central part of the optical potential used in
this work for different energies, such that the amplitude of the eikonal
phase corrected nucleon spinors could also be corrected by the
corresponding energy-dependent amplitude modification.

The focusing and eikonal phase correction has an impact on the outgoing
nucleon flux, and both the cross sections calculated in our
approach and the results found in \cite{Jin1} agree in a different,
but acceptable manner with the experimental data. An overall
phenomenological Perrey factor \cite{Perrey} (correcting for the
violation of unitarity in our simplified approach) which applies in
the same manner to the nucleon wave functions irrespective of
the electron wave functions used, only leads to small corrections
to the cross sections of the order of a few percent and
has been neglected. However, even though the
distorted nucleon wave function approach presented in \cite{Jin1}
is more ambitious than the present one, it is unable to reproduce the
experimental data on a much higher level of precision, since it is
still based on a SPS model, which fails to account for many physical
aspects of the inclusive scattering process.
The relevant observation presented in this paper is the fact that Coulomb
corrections turn out to be compatible with the EMA, with no
strong dependence on the nuclear model used. 

It is clear that the same wave functions for the outgoing nucleon have to
be used for the electron PWBA and exact DWBA calculations,
since we are only interested in the role of the electronic part of the Coulomb corrections. As mentioned above, for bound nucleons wave functions
generated from a non-relativistic Woods-Saxon potential and a Coulomb potential
for the protons were used. The Woods-Saxon potential, which
included an LS coupling term, was optimized in
such a way that the experimental binding energies
of the upper proton shells and the rms of the
nuclear charge distribution were reproduced correctly.
The rms radius of the neutron density distribution was taken from a
recent study presented in \cite{Clark},
the binding energies, which also enter in the calculation of the
wave functions of the outgoing protons and in the phase space
factors, were taken from \cite{Batenburg}.

\subsection{Electromagnetic nucleon current}
The nucleon current 
\begin{equation}
j^\mu_N=e\bar{\Psi}_{N,f} {\hat{J}}_N^\mu \Psi_{N,i} \, ,
\end{equation}
was modeled by using
the cc1 current introduced by de Forest, given by the operator \cite{deForest}
\begin{equation}
{\hat{J}}_N^\mu(cc1)=(F_1+\kappa F_2) \gamma^\mu
-\frac{({p_i}^\mu+p_f^\mu)}{2 m_N} \kappa F_2,
\end{equation}
with the so-called Dirac and Pauli form factors $F_{1,2}(q^2)$ related to
the the Sachs form factors $G_E(q^2)$, and $G_M(q^2)$ \cite{Hoehler,Qattan,Castillo},
according to
\begin{equation}
G_E(q^2) \equiv F_1(q^2)+\frac{\kappa q^2}{(2m)^2} F_2(q^2)=
F_1(q^2)-\kappa \tau F_2(q^2) ,
\end{equation}
\begin{equation}
G_M(q^2) \equiv F_1(q^2)+ \kappa F_2(q^2),
\end{equation}
where $q^2= (p_f^\mu - p_i^\mu)^2$, and $\kappa$ is the anomalous magnetic moment
in units of nuclear magnetons ($\kappa_p = \mu_p-1 = 1.792847$ for  the proton, and
$\kappa_n = \mu_n = - 1.913043$  for the neutron). 
We emphasize that $p_i^\mu$, $p_f^\mu$ and correspondingly $q^2$ are differential
operators acting on the initial and final state nucleon wave functions
$\Psi_{N,f}$ and $\Psi_{N,i}$, and should
not be confused naively with corresponding asymptotic C-number momenta of the nucleon.
However, for the sake of convenience, we maintain here this notational ambiguity, which is
irrelevant in the case of free particles described by plane waves.
We note that we used a completely analogous expansion of the $q^2$-dependence of the form
factor operators as it was used by Knoll for the photon propagator in
eqns. (\ref{scalarop},\ref{wif},\ref{approx}) in our numerical calculations.

We adopted the dipole formula for the nucleon form factors according to
\begin{equation}
G_E^p(Q^2=-q^2)=G_D(Q^2)=\Biggl( 1+\frac{Q^2}{0.71 \, \mbox{GeV}^2} \Biggr)^{-2} , \label{form}
\end{equation}
with a magnetic proton form factor given by
$G_M^p(Q^2) = (1 + \kappa_p) G_E^p(Q^2)$, or equivalently
\begin{equation}
F_1^p=\frac{1+\mu_p \tau}{1+\tau} G_E^p, \quad F_2^p=\frac{1}{1+\tau} G_E^p.
\end{equation}
In the case of the neutron
we used the Galster parametrization \cite{Galster} for the (small) electric form factor
\begin{equation}
G_E^n(Q^2)=-\tau \kappa_n G_D(Q^2) \xi_n(Q^2),
\end{equation}
\begin{equation}
\xi_n(Q^2)=\Biggl( 1+\frac{Q^2}{0.63 \, \mbox{GeV}^2} \Biggr)^{-1},
\end{equation}
consistent with the world data \cite{Platchkov90,Donnelly92}, and $G_M^n(Q^2)=\kappa_n G_D(Q^2)$,
such that
\begin{equation}
F_1^n=\frac{G_E^n+\tau G_M^n}{1+\tau}, \quad F_2^n=\frac{G_M^n-G_E^n}{\kappa_n (1+\tau)}.
\end{equation}

An alternative to the cc1 current is the cc2 current
\begin{equation}
{\hat{J}}_N^\mu(cc2)
=F_1 \gamma^\mu + \frac{i \kappa}{2 m_N} \sigma^{\mu \nu} q_\nu F_2 ,
\end{equation}
which was also used in order to check how the ratio of the cross
sections behaves for different models of the nuclear current.
None of the expressions for the current (cc1,cc2)
is fully satisfactory and both expressions fail to fulfill current
conservation, but given the fact that we focus mainly on the electronic
part of the problem, the simple choices given above provide a satisfactory
description of the proton current. An advantage of the cc2 choice is the fact that integration
by parts allows one in a simple way to get rid of the momentum
operators acting on the nucleon wave functions.

The asymptotic momenta of the nucleons were calculated from energy and
momentum conservation, i.e. the energy of the knocked-out nucleons
was reduced by their initial binding energies.
We also performed calculations with enhanced binding energies in order
to take into account that the average nucleon
removal energy is higher than the average binding energy.
This strategy leads to a shift of the quasi-elastic peak to higher
energy transfer, but has no significant impact on the general
conclusions drawn in this paper concerning the applicability of the EMA.
We found in general that different choices for the nucleon current
do not affect significantly the relative behavior of PWBA, EMA and
DWBA calculations. This is not the case for the absolute cross
section themselves.

The accuracy concerning the calculation of cross sections
is limited to about $1\%$ due to the truncation of the Knoll
expansion eq. (\ref{approx}) and the finite resolution of the grid that
has been used for the modeling of the nucleus. 
The numerical evaluation of transition amplitudes was performed by putting
the nucleus on a three dimensional cubic grid with a side length of $30$ fm and
a grid spacing of $(30/n_{grid})$ fm, and convergence was checked by using
different side lengths and grid resolutions.
The number of necessary grid points $\sim n_{grid}^3$ is mainly dictated by
the wave length of the oscillatory behavior of the matrix element eq. (\ref{wif})
$\sim e^{i(\vec{k}_i-\vec{k}_f-\vec{p}_f)}$ and was in the range of $70^3$ to $130^3$
in order to ensure an accuracy better than $10^{-1}$ percent for the values of
the integrals. The accuracy of the solid angle integration of the
$(e,e'N)$ cross section was better than $0.05\%$.
The truncation of the partial wave expansion of the electron wave
functions \cite{Basel3} was performed such that all partial waves
corresponding to angular momenta up to $|\kappa|=70$ were
taken into account, which guaranteed an accurate evaluation of the
electron wave functions and corresponding first and second order
derivatives appearing in the Knoll expansion in the relevant nuclear
vicinity.

\section{Results}
We first discuss some phenomenological properties of the quasi-elastic peak, calculated within the
electron PWBA, EMA, and DWBA  SPS framework, and from an experimental point of view.
Within a simplified picture like the Fermi gas model, the position of the quasi-elastic peak
for $(e,e')$ scattering with initial electron energy $\epsilon_i$ and scattering
angle $\Theta_e$ is expected at
\begin{equation}
\omega_{peak}=\frac{Q^2_{peak}}{2 m_N} +\bar{\epsilon}_{_{rem}}=
\frac{\epsilon_i^2 (1 - \cos \Theta_e )}
{m_N+(1-\cos \Theta_e) \epsilon_i}+\bar{\epsilon}_{_{rem}}, \label{position1}
\end{equation}
where $\bar{\epsilon}_{_{rem}}$ can be interpreted as a phenomenological average removal
energy of approximately $44$ MeV \cite{Whitney74}. These $44$ MeV include a Coulomb
shift of approximately $8$ MeV, such that the physically relevant Coulomb corrected removal
energy is rather given by $\sim 36$ MeV, if the comparably small contribution of neutrons
to the $(e,e')$ cross section is not taken into account.
The naive expression eq. (\ref{position1}) predicts a difference of the position of
the PWBA and EMA quasi-elastic peaks of $7.1$ MeV for $\epsilon_i=485$ MeV and $\Theta_e=60^o$,
and $11.7$ MeV for $\epsilon_i=310$ MeV and $\Theta_e=143^o$.
A fit of the experimentally measured peak positions \cite{Zghiche94,SaclayPeak} leads to  
\begin{equation}
\omega_{peak}^{exp} = \frac{{\epsilon'}_i^2 (1 - \cos \vartheta)+\tilde{m}_N \bar{E}}{\tilde{m}_N +
\epsilon'_i (1-\cos \vartheta)} \label{effslope2}
\end{equation}
with $\tilde{m}_N$ and $\bar{E}$ as fitting parameters, and ${\epsilon'}_i=\epsilon_i-\bar{V}$
is the effective initial electron energy with $\bar{V}=-20 \, \mbox{MeV}$ in the present fit.
Eq. (\ref{effslope2}) incorporates also the background in the experimental data due
to physical mechanisms like correlation effects and meson exchange
currents which are not included in the SPS model.
The experimental data are reproduced
in a very satisfactory manner by eq. (\ref{effslope2})
for $\Theta_e=60^o$ with $\tilde{m}_N=721 \, \mbox{MeV}$
and $\bar{E}=12.8 \, \mbox{MeV}$ in the energy range $\epsilon_i=262...645$ MeV,
and for $\Theta_e=143^o$ with
$\tilde{m}_N=662 \, \mbox{MeV}$ and $\bar{E}=10.0 \, \mbox{MeV}$ in the energy range
$\epsilon_i=140...420$ MeV. From eq. (\ref{effslope2}), one obtains a difference of the position of
the PWBA and EMA quasi-elastic peaks of $8.3$ MeV for $\epsilon_i=485$ MeV and $\Theta_e=60^o$,
and $13.4$ MeV for $\epsilon_i=310$ MeV and $\Theta_e=143^o$.

\begin{figure}[htb]
\begin{center}
  \centerline{
    \mbox{\includegraphics[width=2.80in]{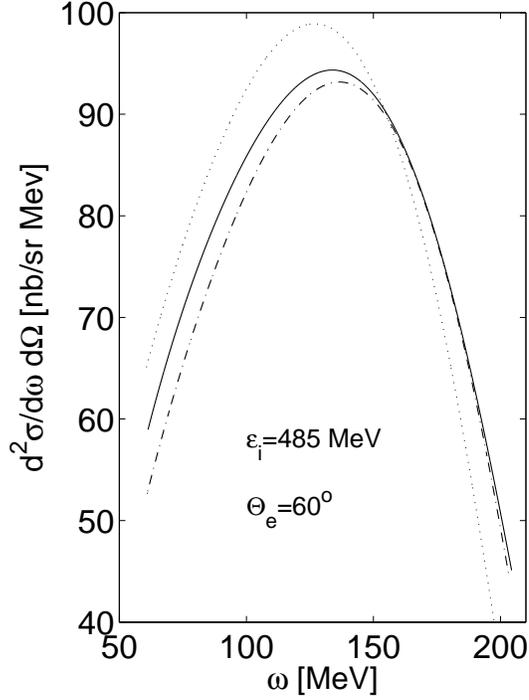}}}
        \caption{The differential cross section for $^{208}$Pb(e,e')
                 scattering for $\epsilon_i=485 \, \mbox{MeV}$
                 and $\Theta_e=60^o$, based on
                 the eikonal approximation for the nuclear current.
                 The dotted line is the electron PWBA result,
                 the dash-dotted line
                 shows the EMA result with an effective potential
                 value $\bar{V}=-19$ MeV,
                 and the solid line displays the DWBA calculation.}
        \label{cross_485_60_ei}
\end{center}
\end{figure}

\begin{figure}[htb]
\begin{center}
  \centerline{
    \mbox{\includegraphics[width=2.82in]{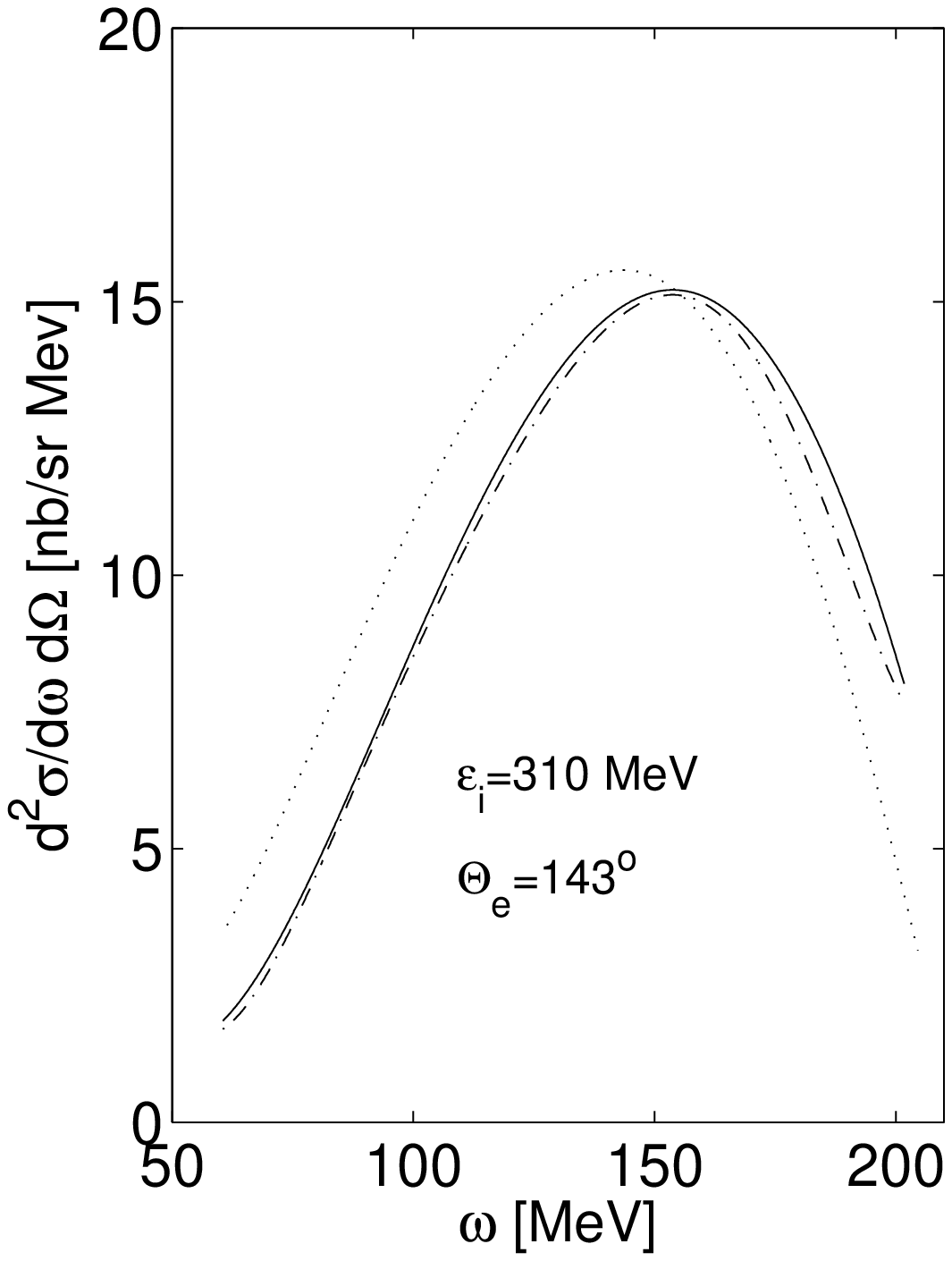}}
  }
        \caption{The results corresponding to Fig. \ref{cross_485_60_ei} for the
                 ($310 \, \mbox{MeV}, \, \Theta_e=143^o$) kinematics.}
        \label{cross_310_143_ei}
\end{center}
\end{figure}

\begin{figure}[htb]
\begin{center}
  \centerline{
    \mbox{\includegraphics[width=2.80in]{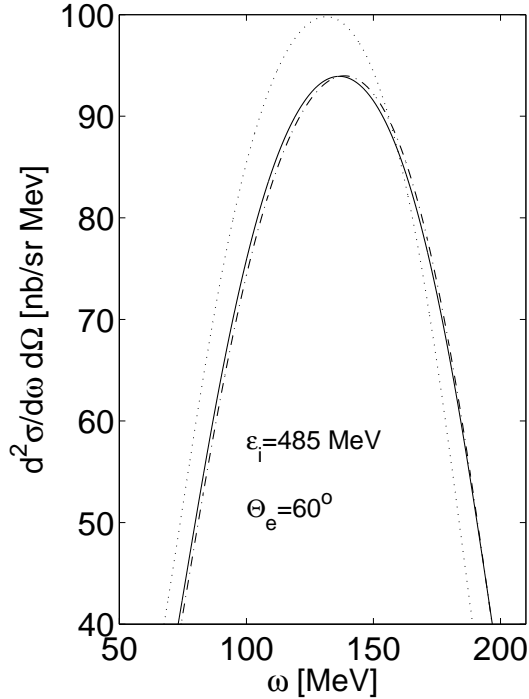}}}
        \caption{The differential cross section for
                 $^{208}$Pb(e,e') scattering
                 at two different electron energies and
                 scattering angles (see also Fig. \ref{cross_485_60_pl}),
                 calculated by using plane wave functions for the final
                 state nucleons (dotted line: PWBA, dash-dotted line:
                 EMA with effective potential $\bar{V}=-19$ MeV,
                 solid line: DWBA). The peaks are slightly narrower
                 than in the nuclear eikonal model, since effects
                 of the final state interactions described by the
                 optical potential are absent.}
        \label{cross_485_60_pl}
\end{center}
\end{figure}

\begin{figure}[htb]
\begin{center}
  \centerline{
    \mbox{\includegraphics[width=2.82in]{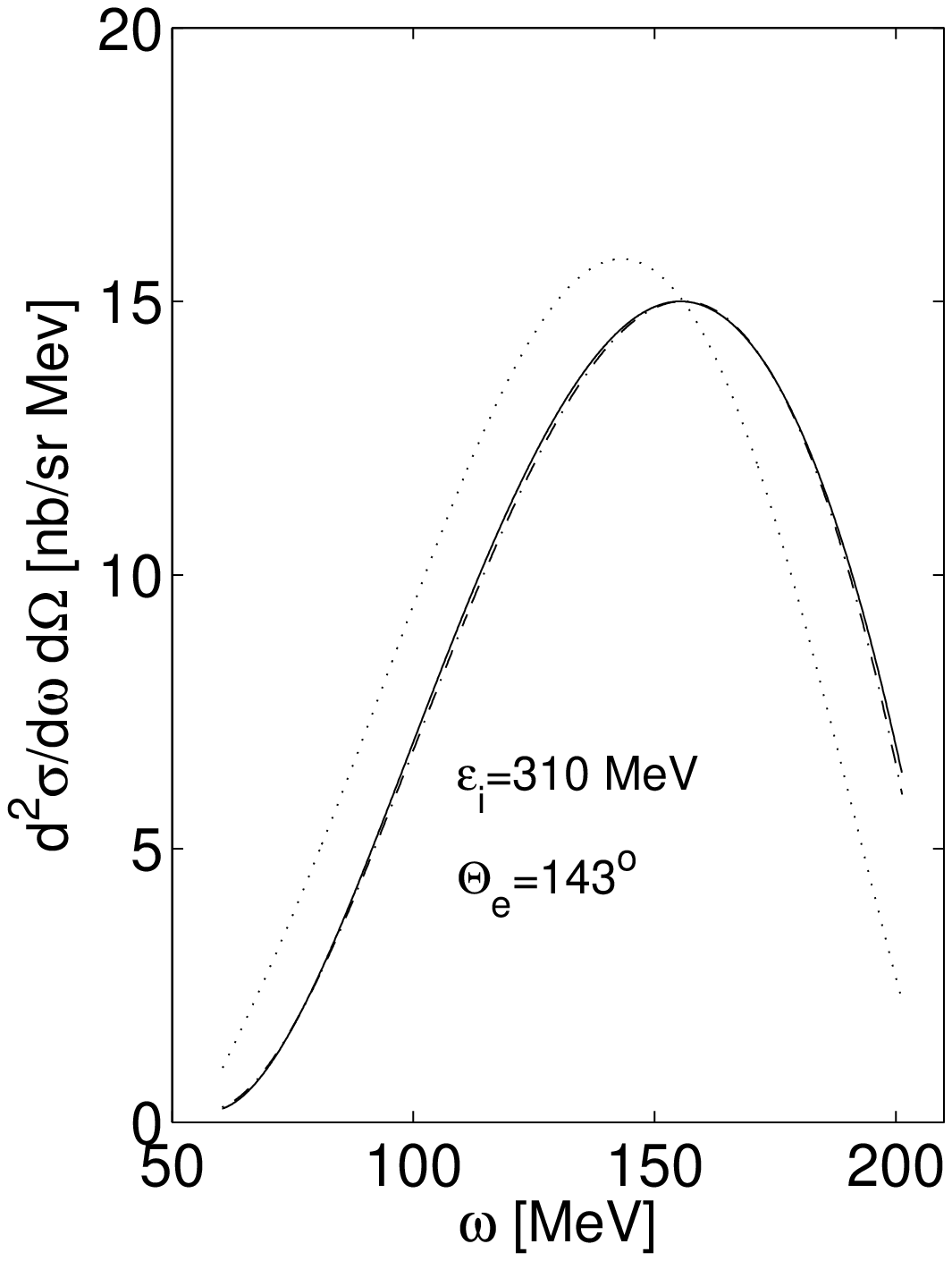}}}
        \caption{The results corresponding to Fig. \ref{cross_485_60_pl}
                 for the
                 ($310 \, \mbox{MeV}, \, \Theta_e=143^o$) kinematics.}
        \label{cross_310_143_pl}
\end{center}
\end{figure}

The corresponding calculations displayed in Figs. \ref{cross_485_60_ei} and \ref{cross_310_143_ei}
(Figs. \ref{cross_485_60_pl} and \ref{cross_310_143_pl})
based on the nuclear eikonal model (plane wave approximation)
confirm the behavior of the quasi-elastic peak position described above in a
satisfactory way.  E.g., the peak shifts are $10.9 \, \mbox{MeV}$ ($12.8 \, \mbox{MeV}$) for the
($310 \, \mbox{MeV}, \, 143^o)$ kinematics and
$9.2 \, \mbox{MeV}$ ($7.8 \, \mbox{MeV}$) for the $(485 \, \mbox{MeV}, \, 60^o)$ kinematics.

\begin{figure}[htb]
\begin{center}
  \centerline{
    \mbox{\includegraphics[width=2.40in]{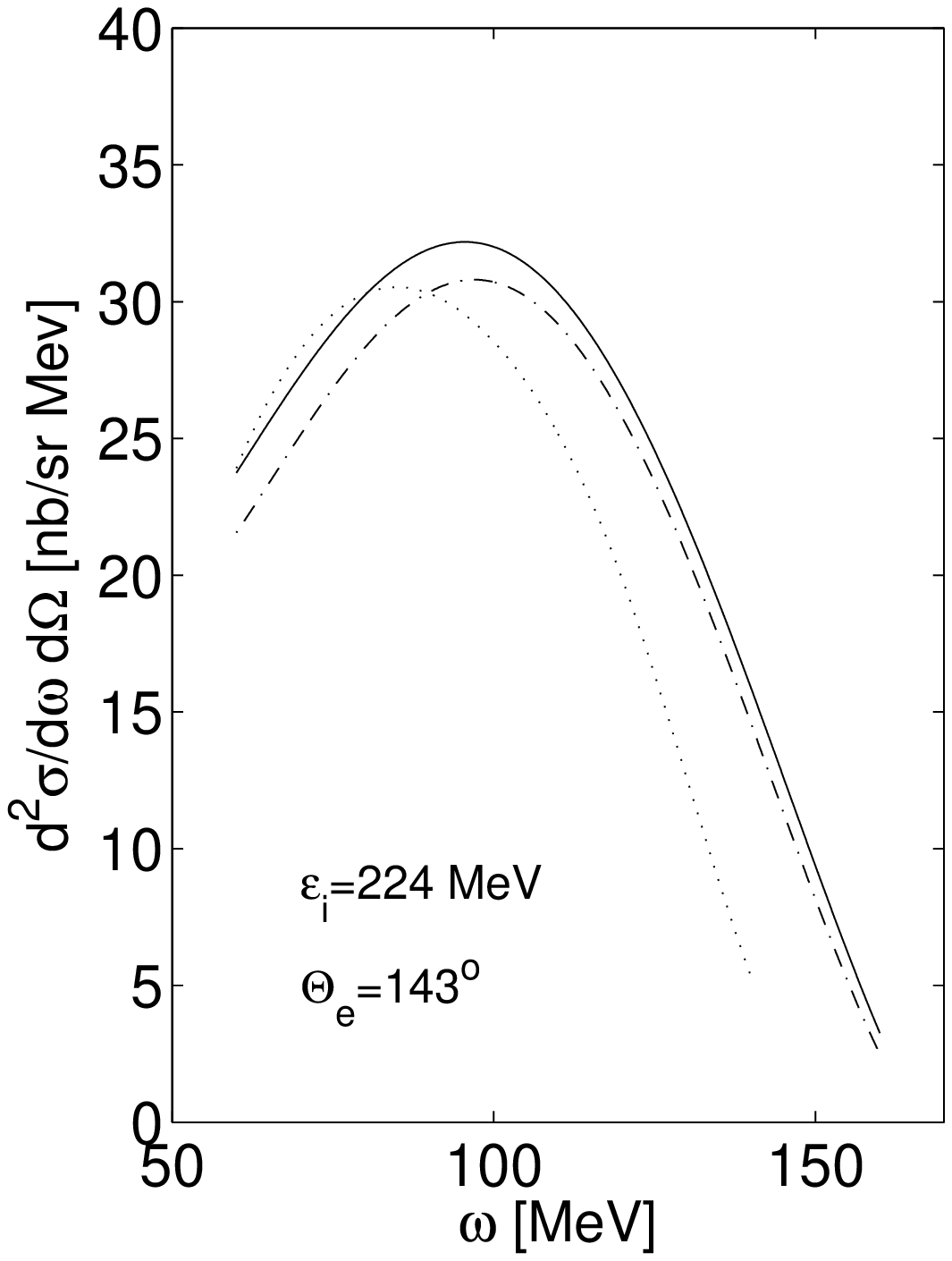}}
  }
        \caption{PWBA, EMA and DWBA cross section for the
                 $(224 \, \mbox{MeV}, \, 143^o)$ kinematics (nucleon eikonal
                 model, $\bar{V}=-19$ MeV). The EMA fails to describe
                 the DWBA cross sections reliably due
                 to the low final state electron energy and the influence
                 of the optical
                 potential on the final state nucleons for low
                 energy transfer.}
        \label{cross_224_143}
\end{center}
\end{figure}

A further comment concerns the relative amplitude of the PWBA and EMA quasi-elastic peak.
An analysis of the Saclay data shows that the maximal $(e,e')$ peak cross section
$\sigma_{peak}=\frac{d^2 \sigma}{d \omega d \Omega} |_{\omega_{peak}}$
scales like $\sigma_{peak} \sim \epsilon_i^{-2.9}$ in the initial electron energy region
$\epsilon_i \sim 485 \, \mbox{MeV}$ for $\Theta_e=60^o$.
Since the EMA cross section can be obtained by multiplying the PWBA cross section, obtained
from effective kinematic quantities, with a focusing factor $[(\epsilon_i-\bar{V})/\epsilon_i]^2$,
a ratio of $\sigma_{peak}^{PWBA}/\sigma_{peak}^{EMA} \simeq (504/485)^{0.9}=1.035$
is expected for the PWBA and EMA peak values,
if the theoretical model is in acceptable accordance with the properties of the physical nucleus.
For $\Theta_e=143^o$, the scaling behavior is $\sim \epsilon_i^{-2.45}$ in the energy
region $\epsilon_i \sim 310 \, \mbox{MeV}$, predicting an approximate ratio
$\sigma_{peak}^{PWBA}/\sigma_{peak}^{EMA} \simeq (329/310)^{0.45}=1.027$,
and $\sim \epsilon_i^{-1.9}$ in the energy region $\epsilon_i \sim 224 \, \mbox{MeV}$,
such that the peak values of the PWBA and EMA cross section are nearly identical,
as observed in Fig. \ref{cross_224_143}. We note that the problematic `staircase-like' behavior
of the Saclay data, discussed in \cite{Jourdan96a}, has only a minor influence on the valuations
presented above.

The relative amplitudes and the peak shifts presented in \cite{Jin1} do not exhibit the
behavior expected from the discussion above in a distinct manner, even if one takes into
account that a central potential value $V(0) \simeq -25 \, \mbox{MeV}$ has been used in \cite{Jin1}
for the EMA, whereas an average potential value $\bar{V} = -19 \, \mbox{MeV}$ has been
used for our calculations presented in Figs. 4-8.
Especially, a pronounced enhancement of the DWBA cross sections with respect to the EMA
cross sections in \cite{Jin1} is absent in the present calculations, except for the case where
the final state electron energy is specifically low, as shown in Fig. \ref{cross_224_143} below.

For the $(485 \, \mbox{MeV}, \, 60^o)$ kinematics,
a slight shift to the left of the DWBA curve with respect to the EMA
curve is observed. Still, an EMA-type behavior of the DWBA result
can be noticed.
For lower energy transfer, the influence of the optical potential
on the nucleons with relatively low energy is stronger and becomes
less important at higher energy transfer, where the EMA and DWBA
match almost perfectly. The EMA/DWBA mismatch
is virtually absent in the case where plane waves are used for
the outgoing nucleon wave functions (see Figs. \ref{cross_485_60_pl}
and \ref{cross_310_143_pl}), i.e. when the distortion of the final
state nucleons is not taken into account.
From a practical point of view, one may observe that the EMA and the
DWBA curve match almost perfectly in the energy range
$80 \, \mbox{MeV} < \epsilon_i < 180 \, \mbox{MeV}$,
if an effective potential value of $-16$ MeV is used and the EMA
results are additionally amplified by $2\%$. However, from the physical
point of view, one should argue that an effective potential value of
$\bar{V}=-19 \, \mbox{MeV}$ is an optimal choice,
if the energy transfer and the corresponding energy of the final
state nucleons is high enough such that the impact of the final
state interaction on the cross  described by the optical potential
becomes less important. The DWBA cross sections are
still larger by $1-2 \%$ than the EMA cross sections in this
kinematical region, i.e., a slight `overfocusing' can be observed.
Still, one should keep in mind that the accuracy of the present
calculations is limited to a similar order of magnitude
by the use of the expansion eq. (\ref{approx}).
For the $(310 \, \mbox{MeV}, \, 143^o)$ kinematics, the situation is
similar, although the impact of the final state interaction described by the
optical potential in the low energy transfer region to the left of the quasi-elastic peak
is less pronounced. This is also due to the fact that a marginally
larger effective potential of $20 \, \mbox{MeV}$ would be appropriate for this kinematics,
leading to a better fit of the EMA and DWBA for higher energy transfer. Furthermore,
the quasi-elastic peak is also shifted to higher energy transfer compared to the
$(485 \, \mbox{MeV}, \, 60^o)$ kinematics, additionally reducing the influence of the
optical potential.

As expected, in the case of the $(224 \, \mbox{MeV}, \, 143^o)$ kinematics
displayed in Fig. \ref{cross_224_143}, the EMA starts to fail.
The reason is twofold. Firstly, the energy of the final state electrons
is relatively small, and the focusing effect becomes more important \cite{Aste2007}, such that
the DWBA cross section becomes significantly larger than the EMA prediction.
Secondly, the quasi-elastic peak is shifted towards smaller energy transfer,
where the optical potential increasingly distorts the wave functions of the
final state nucleons. The momentum transfer squared is still relatively
large, and is therefore not a contributing factor to the failure of the
EMA. Note also that the shifts of the EMA and DWBA peaks with respect
to the PWBA peak are still of comparable size. Again, in order to
describe the DWBA result within an EMA framework, one could increase
the effective potential to a higher value such that the EMA and
DWBA curve match better in the region of large energy transfer.
An other strategy would be to derive an optimized effective potential
which shifts the maximum of the EMA and DWBA peak to the same position
and to renormalize the EMA curve
by an appropriate factor subsequently.

As a general remark, one may observe that the Coulomb corrections
introduced within the framework of the asymptotic
expansion eq. (\ref{approx}) according to Knoll depend on derivatives
of the electron charge and current densities, or, by partial integration,
on the nuclear four-current. Since the corresponding nuclear response
is rather smooth (differently from elastic and transition form factors
to discrete levels), and only leading terms in the expansion are relevant,
one could expect that the dependence of Coulomb effects on different
models of the nuclear current is rather weak.

We finally comment on the electron eikonal calculations (EDWBA) presented
in \cite{Basel1}, which originally seemed to be compatible with the
Ohio group results in \cite{Jin1}.
The PWBA and EDWBA curves in Fig. 3 of \cite{Basel1} show a indeed
similar behavior as the  PWBA and DWBA results in \cite{Jin1}.
\begin{figure}[htb]
\begin{center}
        \includegraphics[width=8.0cm]{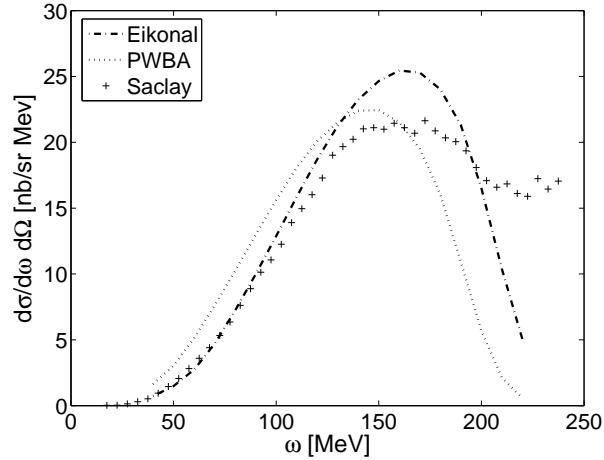}
        \caption{Cross sections for $(e,e')$ scattering with
           $\epsilon_i=310 \, \mbox{MeV}$
           and $\vartheta=143^o$ obtained in \cite{Basel1}.
           The plots displays a similar behavior
           as the results presented in \cite{Jin1}, however,
           the Coulomb corrected eikonal cross
           sections are too large due an overestimation of
           the electron wave function focusing. A slightly different
           nuclear model was used than in the present work, and the
           reduced strength due to correlated nucleons had been
           neglected.}
        \label{eiko143}
\end{center}
\end{figure}
As in \cite{Jin1}, the Coulomb corrected cross section is larger at
the peak than in the PWBA case, seemingly in contradiction with the EMA.
A typical example from \cite{Basel1} is displayed in Fig. \ref{eiko143},
where also experimental data taken at Saclay have been included
\cite{Zghiche94}.
However, there is a simple explanation for this discrepancy.
For the EDWBA calculations in \cite{Basel1}, a focusing value for
the electron cross section was used which corresponds basically to the
central value of the electrostatic potential, i.e. $V(0)=25 \, \mbox{MeV}$,
which leads to an overestimation of the Coulomb corrected cross sections.
Note that the average momentum of the electrons is well described by
the eikonal integral and also corresponds also to an average value of
approximately $19 \, \mbox{MeV}$.

\begin{figure}[htb]
\begin{center}
        \includegraphics[width=8.0cm]{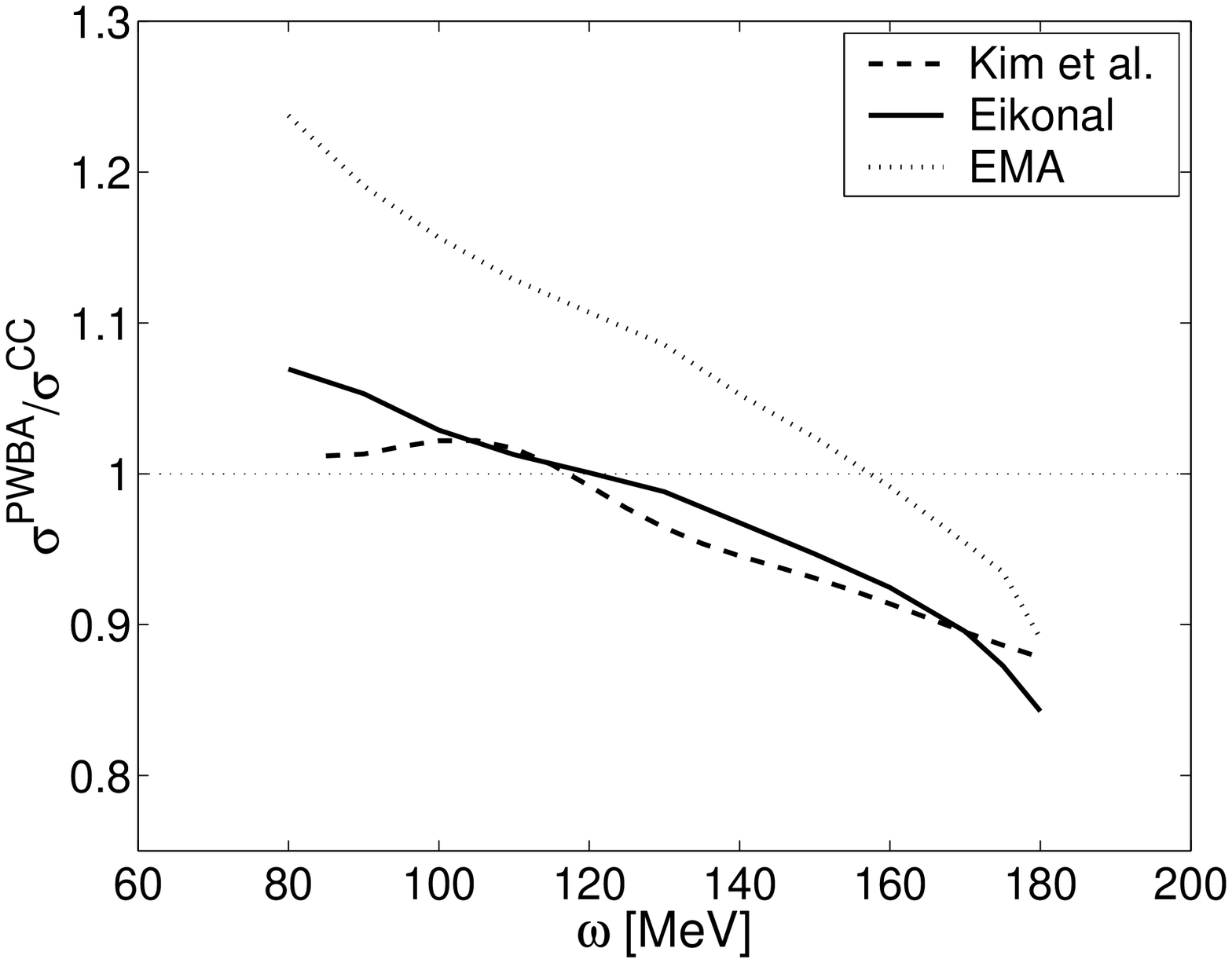}
        \caption{Comparison of Coulomb corrections
                ($\epsilon_i=485 \, \mbox{MeV}$,
                 $\vartheta=60^o$) for different approaches in
                 \cite{Basel1}. For the EMA,
                 an effective potential value of $25 \, \mbox{MeV}$
                 was used.}
        \label{ratio_60}
\end{center}
\end{figure}

\begin{figure}[htb]
\begin{center}
        \includegraphics[width=8.0cm]{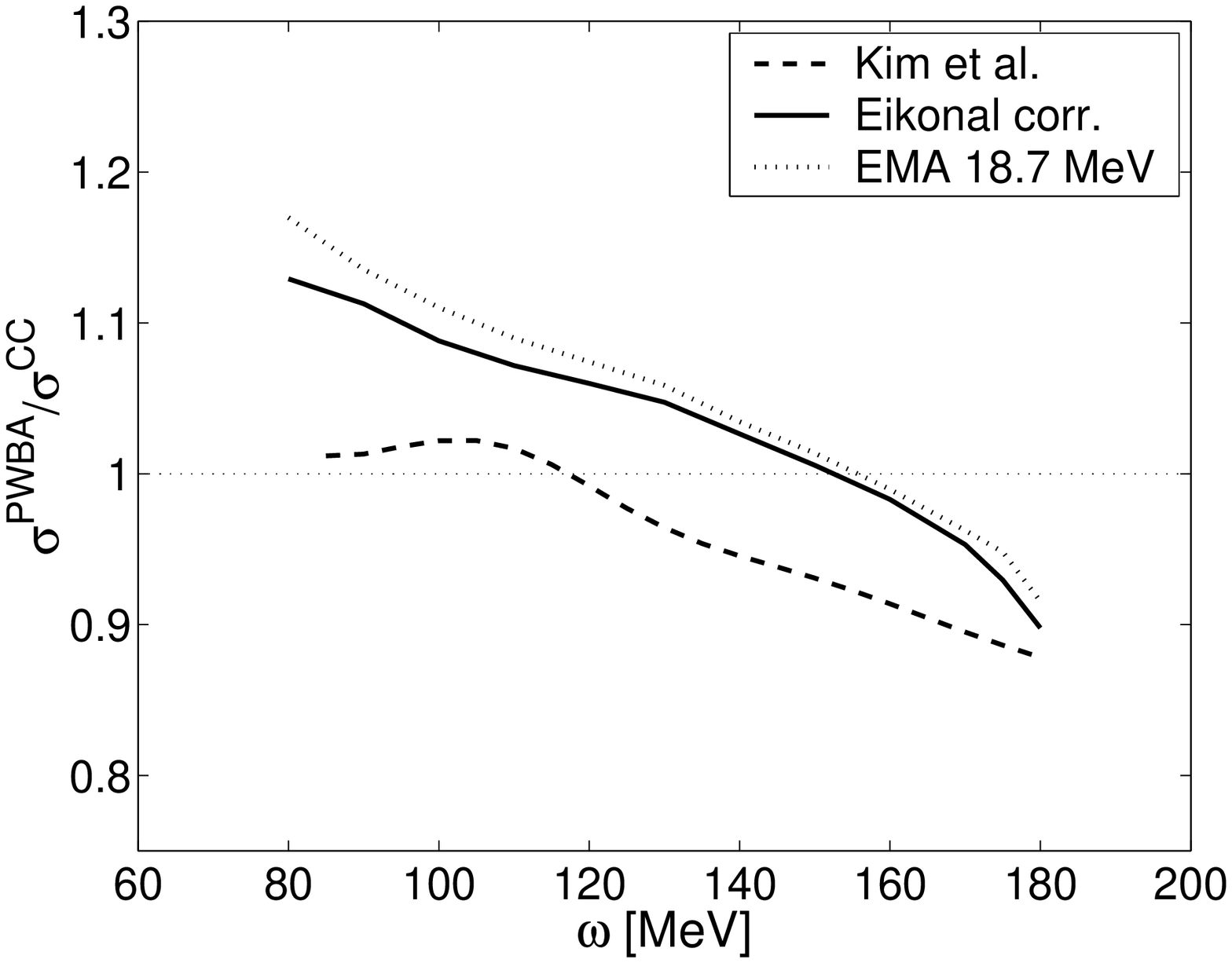}
        \caption{The same as in Fig. \ref{ratio_60}, but with
                 corrected focusing in the eikonal calculation
                 and an EMA curve obtained by using an average
                 potential $\bar{V}=18.7 \, \mbox{MeV}$.}
        \label{ratio_60corr}
\end{center}
\end{figure}

Therefore, the results presented in \cite{Basel1} can be be corrected
by implementing the focusing of the electron wave functions obtained
from the exact electron wave functions in the program originally used
in \cite{Basel1}, with the result illustrated below for Fig. 4
in \cite{Basel1}, which is displayed
again as Fig. \ref{ratio_60} in this paper.
The figure shows the ratio of the cross sections,
calculated in PWBA, with the Coulomb corrected cross sections according
to the EDWBA and EMA. Firstly, the EMA curve has to be adapted
to an effective potential of $19 \, \mbox{MeV}$ instead of $25 \,
\mbox{MeV}$.
This slightly reduces the ratio $\sigma_{PWBA}/\sigma_{EMA}$ and
moves the corresponding dotted curve closer to one (the horizontal line).
Secondly, the focusing factors of the EDWBA calculations must be corrected.
E.g., for $\epsilon_i=485 \, \mbox{MeV}$
and $\epsilon_f=385 \, \mbox{MeV}$, the total focusing factor
entering the EDWBA for the initial and final state electron originally
used in \cite{Basel1} corresponded approximately to the numerical value
\begin{equation}
\frac{(485+25)^2 \times (385+25)^2}{485^2 \times 385^2} \simeq 1.254. \label{example1}
\end{equation}
The correct total focusing factor should rather be
\begin{equation}
\frac{(485+19)^2 \times (385+19)^2}{485^2 \times 385^2} \simeq 1.189. \label{example2}
\end{equation}
Accordingly, the EDWBA cross section has to be reduced by
$5.2 \%$ at $\omega=100 \, \mbox{MeV}$
and the corresponding solid curve moves upwards in the plot.
The consequences are shown in Fig. \ref{ratio_60corr}. Note that for Fig. \ref{ratio_60corr}, also locally varying focusing factors obtained
from exact solutions of the Dirac equation were used in conjunction
with the eikonal approximation for the phase of the electron wave
functions. An attempt to calculate corrections to the focusing
at some distance to the nuclear center has already been presented
in \cite{Knoll}, which, however, does not lead to reliable predictions
in the important surface region of the nucleus and resulted in the
overestimation of the focusing in \cite{Basel1}.
For the EMA calculation presented in Fig. \ref{ratio_60corr}, a
slightly smaller value $\bar{V}=18.7 \, \mbox{MeV}$ than in the
calculational example above was used, since this value has been
determined experimentally to be $\bar{V}=18.7 \pm 1.5 \, \mbox{MeV}$
for $^{208}$Pb \cite{Gueye99} by a comparison of inclusive
electron and positron scattering data. This independent experimental
finding is remarkable, since it also supports that an effective
potential which is slightly smaller than the naive mean potential
$\bar{V} \sim 4 V(0)/5 \sim 20$ MeV is adequate. One might speculate that
the contribution of nucleons near the surface of the nucleus
to inclusive scattering is enhanced, leading to the observed reduction of
the effective potential. 

\section{Conclusions}
As a basic result of this work we conclude that the EMA with an
effective potential $\hat{V}=-19 \, \mbox{MeV}$ is a valid approximation
for the description of Coulomb distortions in the kinematic region where
the momentum transfer squared is larger than $(300 \, \mbox{MeV})^2$
(such that the length scale of the exchanged photon is smaller than
the typical size of a nucleus), the energy of the scattered electron
is larger than $150 \, \mbox{MeV}$ (such that the semiclassical
description of the electron wave functions in the nuclear vicinity
is valid) and the energy transfer $\omega=\epsilon_i-\epsilon_f$ is
larger than $\sim 140 \, \mbox{MeV}$
(such that the distortion of the final state nucleon wave functions
is moderate). In the case mentioned above, and for a typical energy
range where the initial electron energy is of the order of some hundreds
of MeV, the DWBA cross sections are generally larger by $1-2 \%$ than
the EMA cross sections. 
If the energy transfer $\omega$ is smaller than $140 \, \mbox{MeV}$,
the DWBA cross section can still be approximated by
an EMA calculation with a phenomenological effective potential $\hat{V}$ and
a minor amplitude correction. However, in such a case one has to rely on the
model used to describe the nuclear current.

Other kinematic situations than those presented in this paper have been investigated,
which showed that the behavior of the forward and backward
scattering angle kinematics presented in this paper is typical.

The coincidence of EMA and DWBA cross sections is rather impressive.
As a test, the PWBA and EMA cross section calculations which are based
on plane Dirac waves for electrons were evaluated by both
using Dirac plane waves and by performing a limit $Z \rightarrow 0$ in the
full Coulomb solver for the electron wave functions.
Performing such a limit in the DWBA solver leads to the same
cross sections as obtained from the plane wave calculations,
such that the observed coincidence served as a numerical test
for the two unrelated calculational techniques.

It is instructive to compare Fig. \ref{cross_485_60_ei} in this work
to Fig. 16 in the classic work by Rosenfelder on quasi-elastic
electron scattering on nuclei \cite{Rosenfelder80}.
It should be noted that Rosenfelder explicitly mentioned already in
his work that the effective potential $\bar{V}$ is close to the
\emph{mean value} of the electrostatic potential of the nucleus.
However, he then wrote down the explicit expression for the
\emph{central} value
\begin{equation}
V(0)=-\frac{3 Z \alpha}{2R}
\end{equation}
of the potential of a homogeneously charged sphere with radius $R$,
which is related to the corresponding mean value by
$\bar{V} \simeq 4 V(0)/5$.
This minor, but not irrelevant misapprehension for the case of the EMA,
has propagated in the literature since then (see, e.g., \cite{Hotta84}).
Apart from the fact that the initial electron energy is marginally larger and
the effective potential used by Rosenfelder is given by $\bar{V}=-24.9 \, \mbox{MeV}$,
Fig. 16 in \cite{Rosenfelder80} can directly be related to the result presented in
Fig. \ref{cross_485_60_ei}.

Despite many theoretical and experimental efforts to clarify
to what extent a suppression of the Coulomb sum rule exists, clear
answers to the problem are still missing. The origin of the
discrepancy between the results presented in this paper and the findings
of the Ohio group remains hitherto unclear, but one may speculate that
it could be related to the relativistic nucleon potential model
used by the Ohio group, which may influence the resulting nuclear current
(spinor distortion or negative energy contributions) in an unexpected way.
An overview on different theoretical attempts to implement
Coulomb corrections in the analysis of experimental data
can be found in \cite{KimSum}. According to a reanalysis of
experimental data based on the validity of the EMA, which is
supported by the present work, Meziani and Morgenstern claimed
that a suppression of the longitudinal structure function of
about $40$\% exists at the effective momentum transfer of
$500$ MeV/c, and tried to explain the suppression by a change
of the nucleon properties inside the nuclear medium \cite{MorgenMeziani}.
However, in view of the fact that theoretical questions about the
interpretation of relativistic nucleon potential models persist and further
related effects like meson exchange currents and correlations
have not been taken into account in a satisfactory manner in
theoretical Coulomb distortion calculations up to the present, and since
the quality of experimental data may be questionable in some cases
(see, e.g., \cite{Jourdan96a}), it is advisable to await for the
expected experimental TJNAF results at the high momentum transfer region,
in which the relevant correlations become small, whereas Coulomb
corrections become crucial to extract information about the structure
functions and to understand the Coulomb sum rule.

\section*{Appendix A}

The nuclear charge distribution and nucleon optical potentials of heavy nuclei are often
approximated by the help of a Woods-Saxon distribution.
Normalizations and moments of Woods-Saxon distributions are rarely found in the literature,
however, they can be expressed exactly in terms of polylogarithms.

The volume integral over the Woods-Saxon distribution with a range
${\tilde{r}}$ and diffusivity $a$
\begin{equation}
\rho (r)=\frac{\rho_0}{1+e^{(r-{\tilde{r}})/a}}
\end{equation}
can be written with help of the trilogarithm
\begin{equation}
4 \pi \int \limits_{0}^{\infty}
r^2 \rho(r) dr = -8 \pi a^3 \rho_0 {\mbox{Li}}_3 (-e^{{\tilde{r}}/a})
\rightarrow \frac{4}{3} \pi
{\tilde{r}}^3 \rho_0 \, (a \rightarrow 0),
\end{equation}
where the polylogarithms ${\mbox{Li}}_n$ are defined for $|z| <1$
by
\begin{equation}
{\mbox{Li}}_n(z)=\sum \limits_{k=1}^{\infty} 
\frac{z^k}{k^n}.
\end{equation}
For $|z|>1$, the analytic continuation of the
dilogarithm ${\mbox{Li}}_2(z)$ and the trilogarithm ${\mbox{Li}}_3(z)$
can be obtained via the functional equations
\begin{equation}
{\mbox{Li}}_2(z)=-{\mbox{Li}}_2(1/z)-\frac{1}{2} \log(-z)^2-
\frac{\pi^2}{6},
\end{equation}
\begin{equation}
{\mbox{Li}}_3(z)={\mbox{Li}}_3(1/z)-\frac{1}{6} \log(-z)^3
-\frac{1}{6} \pi^2 \log(-z).
\end{equation}
For a nucleus with charge number $Z$ and total
charge $eZ$ one has
\begin{equation}
\rho_0=- \frac{e Z}{8 \pi a^3 {\mbox{Li}}_3 (e^{-{\tilde{r}}/a})}.
\end{equation}
The potential energy of an electron in the center of the nucleus
is given by
\begin{displaymath}
v_0=- e \int \limits_{0}^{\infty} \rho(r) r dr =
-e a^2 \rho_0 {\mbox{Li}}_2 (-e^{{\tilde{r}}/a})
\end{displaymath}
\begin{equation}
= \frac{\alpha Z}{2a} \frac{{\mbox{Li}}_2 (-e^{{\tilde{r}}/a})}
{{\mbox{Li}}_3 (-e^{{\tilde{r}}/a})},
\end{equation}
and the rms radius of the distribution is given by
\begin{equation}
\langle r^2 \rangle = 12 a^2 \frac{{\mbox{Li}}_5 (-e^{{\tilde{r}}/a})}
{{\mbox{Li}}_3 (-e^{{\tilde{r}}/a})}.
\end{equation}
This follows also from the general expression for the moments
of the Woods-Saxon distribution
\begin{equation}
\frac{1}{n!} \int \limits_{0}^{\infty} \frac{r^n}{1+e^{(r-{\tilde{r}})/a}}
dr =-a^{n+1} {\mbox{Li}}_{n+1} (-e^{{\tilde{r}}/a}).
\end{equation}
Note that the corresponding expressions for a homogeneously charged
sphere can be obtained from the limit $a \rightarrow 0$ and
\begin{equation}
{\mbox{Li}}_n(-e^{x}) \simeq -\frac{x^n}{n!}
\quad \mbox{for} \quad x > n^2.
\end{equation}
Choosing the typical parameters ${\tilde{r}}=6.6 \, $fm and
diffusivity $a=0.545 \,$fm for the electric charge distribution
of a $^{208}$Pb nucleus with mass number $A=208$ and
charge number $Z=82$, one finds that these values are compatible with
an rms charge radius of $5.50 \, \mbox{fm}$ and a central Coulomb potential
of $V_0=-25.71 \, \mbox{MeV}$.

\end{document}